# Rethinking Cybersecurity Ontology Classification and Evaluation: Towards a Credibility-Centered Framework

Antoine Leblanc[1[0009-0008-7474-0382]], Jacques Robin[2[0000-0001-7425-2639]], Nourhène Ben Rabah[1[0000-0002-3051-7241]], Zequan Huang[1,2[0009-0001-9551-7621]], Bénédicte Le Grand[1[0000-0002-3813-4093]]

[1] Center for Research in Informatics (CRI), University Paris 1 Panthéon-Sorbonne
[2] Engineering School of Informatics, Electronics and Automation (ESIEA), Paris
{antoine.leblanc, nourhene.ben-rabah, benedicte.le-grand}@univ-paris1.fr, {jacques.robin, zequan.huang}@esiea.fr

**Abstract.** This paper analyzes the proliferation of cybersecurity ontologies, arguing that this surge cannot be explained solely by technical shortcomings related to quality, but also by a **credibility deficit** - a lack of trust, endorsement, and adoption by users. This conclusion is based on our first contribution, which is a state-of-the-art review and categorization of cybersecurity ontologies using the Framework for Ontologies Classification (F4OC) framework. To address this gap, we propose a revised framework for assessing credibility, introducing indicators such as **institutional support**, **academic recognition**, **day-to-day practitioners' validation**, and **industrial adoption**. Based on these new credibility indicators, we construct a classification scheme designed to guide the selection of ontologies that are relevant to specific security needs. We then apply this framework to a concrete use case: the Franco-Luxembourgish research project **ANCILE**[1], which illustrates how a credibility-aware evaluation can reshape ontology selection for operational contexts.

**Keywords:** Ontology Classification, Credibility, Cybersecurity, Quality Assessment.

## 1 Introduction

In the cybersecurity landscape, ontologies have become the semantic backbone of knowledge bases used by reasoning engines: they structure the representation of vulnerabilities, ground alerts and detections in a shared vocabulary, and allow ***Security Orchestration Automation and Response (SOAR)*** platforms to reason about adversarial tactics and appropriate responses. Without them, interoperability between tools, the traceability of automated decisions, and even the simple comparison of incident reports become fragile and error prone. In short, a well-designed ontology is no longer an academic luxury; it is a critical component of modern cyber hygiene.

---

[1] https://www.linkedin.com/company/autonomic-cybersecurity-with-adversarial-learning-and-explanations

This very importance has triggered an overwhelming proliferation. Since 2014, around a hundred different security ontologies have been published. Figure 1 shows this increasing curve, based on a review of ontologies in cybersecurity domain which is the first contribution of our work.

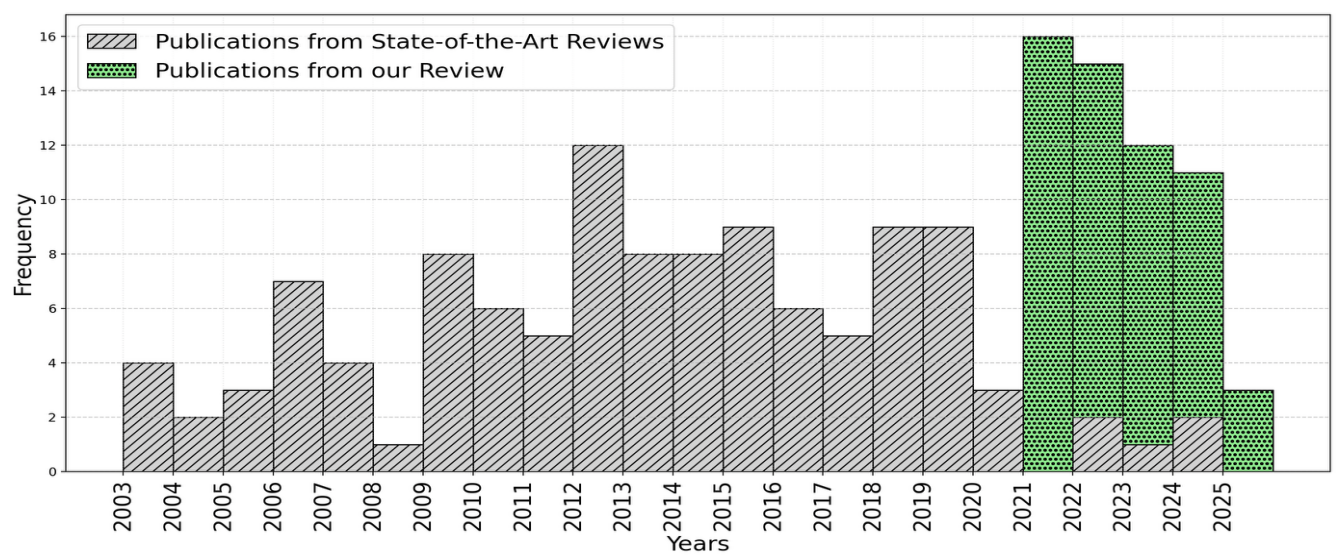

**Fig1.** Number of Published Cybersecurity Ontologies per Year (January 2003–April 2025)

This abundance reveals a lack of reusability, even though this principle is essential for managing data and digital resources according to the ***FAIR principles (Findable, Accessible, Interoperable, and Reusable)*** [1], as argued by Oliveira et al. [2]. This problem of ontologies' abundance has led to our first research question:

- **RQ1.** Why is the reuse of ontologies limited in the cybersecurity domain?

To address this question, we conduct a systematic review of existing cybersecurity ontologies and analyze each one through a structured categorization framework that brings its key properties to light.

Let's now turn to another challenge faced when a cybersecurity team must choose an ontology. They are confronted with a multitude of ontologies in the form of Web Ontology Language (OWL), Neo4j and Prolog implementations, each claiming to cover MITRE ATT&CK™ (a taxonomy of attacker behaviors) [3], MITRE D3FEND™ [4] (a counter-measure ontology mapping how to block or limit ATT&CK tactics)[2], or ISO 27005 [5] – according to Enterprise Strategy Group research, 89% of organizations use MITRE ATT&CK to reduce risk for security operations use-case [6]. However, very few of them are validated by external cybersecurity day-to-day practitioners or widely adopted by operators of ***Security Operations Centers (SOC).*** To guide their choice, teams often rely on intrinsic quality criteria—logical soundness, structural consistency, and syntactic correctness—while overlooking extrinsic properties such as whether the ontology adheres to established standards or how frequently it has been updated. Most of the existing literature has focused on the characterization of ontologies, while few studies have addressed the evaluation of ontology quality in the cybersecurity domain. Moreover, such evaluations remain limited because of the challenges of evaluating their criteria. This led us to formulate a second research question:

- **RQ2.** What additional criteria could be used to characterize cybersecurity ontologies, and how could they be evaluated?

[2] While MITRE ATT&CK was initially published as a mere taxonomy, the more recent MITRE D3FEND which is a genuine OWL ontology includes the ATT&CK taxonomy. In what follows, when we refer to MITRE ATT&CK, we mean the sub-ontology of D3FEND that reuses the ATT&CK taxonomy and elaborate it.

To address this gap, this paper investigates a key complementary criterion: ***Credibility***; and introduce a set of credibility-oriented indicators which constitute our second contribution. Those indicators are alignment with industry standards, validation by independent day-to-day practitioners, demonstrable adoption in operational tools or institutional settings, and long-term maintenance.

Building on this, we propose a new classification framework for cybersecurity ontologies, our *third contribution*, integrating these credibility metrics with previously proposed criteria such as [7]: *Level of application*, Level *of generality*, *Formal expressiveness*, *Logical density* (referred as axiomatization level in [7]) These criteria will be further explained in the background section.

We apply this framework to a concrete use case: selecting an ontology to support ANCILE, a next-generation SOAR framework combining autonomic computing with probabilistic attack detection, attack plan recognition, and automated mitigation, that is being designed in an eponymous Franco-Luxembourgish research project. This application illustrates how the addition of the "credibility" axis reorders the ranking of candidate ontologies and excludes superficially attractive ontologies which lack in fairness and adoption by daily cybersecurity practitioners.

This paper is structured as follows. Section 2 reviews the previous work in classifying ontologies and evaluating their quality; Section 3 reviews cybersecurity ontologies and applies a selected framework to those; Section 4 explores the concept of credibility, and proposes a refined definition and evaluation metrics tailored to this context; Section 5 introduces a new classification framework that incorporates credibility as a core dimension and applies it to a selection of security ontologies; Section 6 presents a concrete application of this framework in the context of the ANCILE project, illustrating how credibility-aware selection impacts practical cybersecurity ontology selection; Section 7 concludes the paper and outlines directions for future research.

## 2 Background Knowledge

### 2.1 Previous work in classifying cybersecurity ontology

To navigate the growing number of cybersecurity ontologies, it is essential to classify and compare them using a consistent and structured approach. One of the most comprehensive efforts in this direction is **F4OC** proposed by Martins et al. [7]. Their objective is to support the engineering, selection, and integration of ontologies (particularly in domains like cybersecurity) by combining conceptual modeling theory with practical needs from software and knowledge engineering. They propose a five-step, orthogonal pipeline that aligns ontology with practical engineering concerns.

- ***Step 1***. Systematic mapping of the state of the art.
- ***Step 2***. Placement of candidate ontology along Guizzardi's ***level-of-application*** axis [8], defining those as *reference* or *operational* or both. A reference ontology is a well-structured representation of a part of reality, designed for knowledge exchange and reuse, whereas an operational ontology is its concrete implementation in practice. Throughout the remainder of this paper, the term reference ontology denotes an ontology that either is a reference ontology itself or is explicitly supported by one.
- ***Step 3***. Application of Guarino's [9] and van Heijst's [10] ***level-of-generality*** taxonomy, categorizing ontologies as *Foundational* (its concepts are broad and apply across multiple domains), *Domain* (its concepts are from a particular domain), *Core* (lies between Foundational and Domain), *Task* (its concepts are specific task within a domain), *Application* (lies between Domain and Task). It also checks if the ontology is well-grounded – i.e. its concepts are inherited from a foundational ontology.
- ***Step 4***. Estimation of *formal expressiveness* that combines Uschold-Gruninger's informal / formal ladder [11] and Giunchiglia et al. lightweight / heavyweight distinction [12].
- ***Step 5.*** Estimation of *logical density*. – referring to the number of concepts and relations.

Another work, from Oliveira et al. [2], assesses security core ontologies against the four FAIR pillars and checks (a) their implementation language, (b) their artifact availability, (c) whether the ontology reuses concepts from some foundational ontology and (d) their concept vocabulary. Their mapping study shows that only a small fraction of the 57 ontologies they examined satisfy all four indicators, especially struggling with Interoperability and public availability.

Meriah and Rabai [13] introduce a domain-focused scheme that distinguishes six main ontologies built to encode security standards from ISO/IEC 27000 [14]. They also provide recommendations for future ontologies based on this standard.

Martins et al.'s F4OC stands out by proposing a more complete categorization framework of ontologies with features that cover essential properties of ontologies: foundational grounding, complexity, operability, and formality. This is why, in the following sections, we propose to describe existing cybersecurity ontologies by analyzing them according to the criteria of this classification framework.

### 2.2 A framework to evaluate ontologies

Following the discussion on how to categorize ontologies, Wilson et al. [15] proposed a recent evaluation framework based on a systematic literature review. It assesses ontology quality along four independent axes:

- The *structural intrinsic* axis focuses on the internal consistency of the ontology itself—how well it follows formal language rules, avoids contradictions, and maintains coherence.
- The *domain intrinsic* dimension examines how accurately the ontology represents its subject matter, typically by comparing it to established standards (e.g., ISO, NIST) and testing it with expert reviews.
- The *domain extrinsic* perspective shifts attention to the ontology's real-world usefulness. Here, it's evaluated as a black box—judged by how well it delivers relevant, accurate, and understandable outputs for users like threat analysts.
- Finally, the *application extrinsic* axis treats the ontology as a software component within a larger technical system. The focus is on technical integration and performance, regardless of domain knowledge.

Among the 19 criteria identified among these axes, ***credibility*** stands out as particularly important – it comes from the domain extrinsic axe. It captures how trusted and widely accepted an ontology is within its community — a key factor in distinguishing impactful ontologies from those with limited practical use. A fuller discussion of credibility is provided in Section 4.

## 3 Survey and characterization of recent cybersecurity ontologies

This section presents our first contribution: a review of cybersecurity ontologies published since 2021, along with their characterization using the F4OC framework.

### 3.1 Review of recent cybersecurity ontologies

We first searched for review papers in cybersecurity ontologies. Along with proposing F4OC, Martins et al. [7] also cataloged thirty-five cybersecurity ontologies. Other authors examine security ontologies, including Oliveria et al. [2], Adach et al. [16], and Rosa et al. [17]. Jarwar [18] concentrates on a specific context, which is IoT-related ontologies. Finally, a few studies do not directly survey ontologies but address closely related knowledge structures like Rahman et al. [19] who review various taxonomies, or Bratsas et al. [20] and Bolton et al. [21] who focus on knowledge graphs.

Overall, these papers cite a total of 114 papers related to 1 security ontologies. However, the 114 security ontologies related papers cited by these articles were all published before 2021 (except for four ontologies from [18]). To get an up-to-date list of cybersecurity ontologies we conducted a review in the spirit of the work of Martins et al. [7] on cybersecurity ontologies published since January 2021. This resulted in adding ***52 new publications of ontologies*** to our list security and cybersecurity ontologies. We used the search string "TITLE-ABS-KEY("Cybersecurity Ontology" OR "Cybersecurity Ontologies")" on the Scopus database we applied the same inclusion and exclusion criteria as [7]. The inclusion criteria were (IC1) Articles that introduce full-scale cybersecurity ontologies, (IC2) Articles that detail specific modules or sub-sets of cybersecurity ontologies. The exclusion criteria (EC1) Articles whose low citation count suggests limited relevance, (EC2) Articles that do not actually put forward an ontology.

Among those, two ontologies stand out due to their foundational grounding: the **Common Ontology of Value and Risk (COVER)** [22]and the **Reference Ontology for Security Engineering (ROSE)** [23]. Both are rigorously grounded in the **Unified Foundational Ontology (UFO)** [24]– both implemented in OntoUML – ensuring clear distinctions between the domain-independent concepts of agents, events, dispositions and situations. COVER proposes a formal ontology of risk, highlighting its deep ties to value. ROSE extends it to security, with some modifications, by making explicit how threats, vulnerabilities, and countermeasures interact in ISO 31000-style risk treatment processes [25].

An OWL ontology **CRATELO** [26] (Cyber Risk And Threat Engineering Linked Ontology) combines a DOLCE-SPRAY [27] foundation, Security Core Ontology (SECCO) and an Ontologies of Secure Cyber Operations (OSCO) to capture the full path from threats and vulnerabilities to countermeasures and mission impact, improving the situation awareness of security operations analysts. Because of its foundation on DOLCE, CRATELO is well-grounded. However, it does not rely on any standards.

**WAVED** [28] (Weakness – ATT&CK – Vulnerability – Engage – D3fend) unifies in a single OWL ontology MITRE ATT&CK tactics, D3FEND counter-measures, ENGAGE [29] engagement activities, the CVE (Common Vulnerabilities and Exposures - a list of disclosed concrete security flaws)[3] and the CWE (Common Weakness Enumeration - recurring software/hardware general weakness classes that are specialized by vulnerabilities)[4] [30] / [31]. Those taxonomies, ontologies and lists are considered de facto standard nowadays because they are openly available, maintained by the trusted MITRE organization, and broadly adopted by tool vendors, government bodies, and practitioners; in the absence of an equivalent rival, they have naturally evolved into the common language of the domain. Thus, WAVED is well rooted into cybersecurity de facto standards, but it lacks foundational grounding.

The Unified Cybersecurity Ontology (UCO) [32] offers a semantic integration of major cybersecurity standards – like Structured Threat Information eXpression (STIX) an open, machine-readable OASIS standard that structures and links cyber-threat intelligence and based on XML [33] – and de facto standards – like CVE – using OWL to support reasoning and interoperability. Unlike STIX, UCO enables advanced queries by linking cyber concepts to general knowledge bases like DBpedia [34] and Yago [35] for instance. However, it lacks foundational grounding.

[3] Example CVE: *CVE-2017-11882* — A vulnerability in Microsoft Word that allows attackers to execute code just by getting a user to open a malicious document

[4] Example CWE: *CWE-89* — SQL Injection: a common weakness where attackers can manipulate database queries to access unauthorized data

### 3.2 Application of Martins et al.'s Framework on post 2021 cybersecurity ontologies for classification purposes

We then applied the F4OC framework to those new 52 ontologies, to see the current trends in security ontology engineering. However, it is difficult to evaluate the size or formalization level of ontologies because the available documents describing them do not provide enough details, as explained by [7]. Therefore, we focused only on the first four criteria provided by F4OC: a) Reference Ontology, b) Operational Ontology, c) Well-groundedness and d) level of generality according to [9] and [10].

A summary of key findings is presented in Table 1, highlighting the percentage of ontologies that satisfy each evaluation criterion. For the "level of generality" dimension, we break down the data into four categories core, domain, application, and task specific. The Foundational category is omitted because it does apply to cybersecurity-specific ontologies.

**Table 1.** Application of the F4OC to the 52 recent publication of cybersecurity ontologies and the 35 coming from [7].[5]

| **Number of Ontology** | **Step 2 (Level of Application)** | | **Step 3 (Level of Generality [7])** | | | | |
|---|---|---|---|---|---|---|---|
| | **Reference Ontology** | **Operational Ontology** | **Well-grounded Ontology** | **According to [9] and [10]** | | | |
| | | | | **Core** | **Domain** | **Application** | **Task** |
| 35 from [7] | 43 % | 83 % | 11 % | 6 % | 77 % | 17 % | 0 % |
| 52 from our review | 25 % | 87 % | 4 % | 0 % | 31 % | 69 % | 0 % |

Our review provides some findings regarding the first research question (**RQ1**) by identifying a key barrier to ontology reuse in cybersecurity: the lack of foundational grounding. Although 87% of the most recent ontologies reviewed claim to serve operational purposes, such as supporting reasoning or inference, only 4% demonstrate grounding in a foundational ontology. This gap has significant consequences. Ontologies developed without a well-established conceptual basis are more prone to inconsistencies, are harder to extend, and are less compatible with broader knowledge ecosystems. A clear example of this is D3FEND, which introduces several useful concepts but lacks ontological rigor, leading to ambiguities and limiting its interoperability [36].

Regarding reuse, we must however nuance our findings because it is not the only aspect to consider. According to [2], Reusability is also assessed by checking if the ontology meets domain-relevant community standard and if it provides rich metadata.

This lack of foundational grounding motivates the next subsection 3.3, where we leverage our characterization framework to propose requirements for selecting a reusable and robust ontology, grounded in both theoretical and operational considerations.

[5] The full results of this analysis are available here https://github.com/AntoineLeblancFr/EdocPAPER.

### 3.3 Selecting a relevant ontology for cybersecurity

While the design of an ontology may be influenced by specific use cases, some properties defined in this framework are broadly desirable and beneficial across various application domains, beyond cybersecurity. First, an ontology which is well-grounded is less prone to error and inconsistencies. Second, a reference ontology has more chance to be reused and respect FAIR principles. Third, an ontology is often built to reason and answer queries, meaning to be operational. Thus, an ontology being well-grounded, being a reference one and being operational is worthwhile to be considered for every project. Filtering the 87 ontologies in our review through those three criteria lenses, *Reference ontology*, *Operational ontology* and *Well-groundedness* singles out CRATELO [26]. Unfortunately, the original artefact appears to be lost; the authors could not retrieve the 2014 code base. We then decided to look for the closest neighbors of CRATELO by relaxing each criterion one by one.

Relaxing the "operational" requirement adds three more candidates, notably COVER [22] and ROSE [23], both grounded in UFO [24] and therefore promising foundations that would, however, still have to be refined with more operational concepts if we wanted an operational ontology on cybersecurity. Dropping the "well-grounded" criterion widens the net to about twenty ontologies and loosening the "reference" criterion yield no additional ontologies.

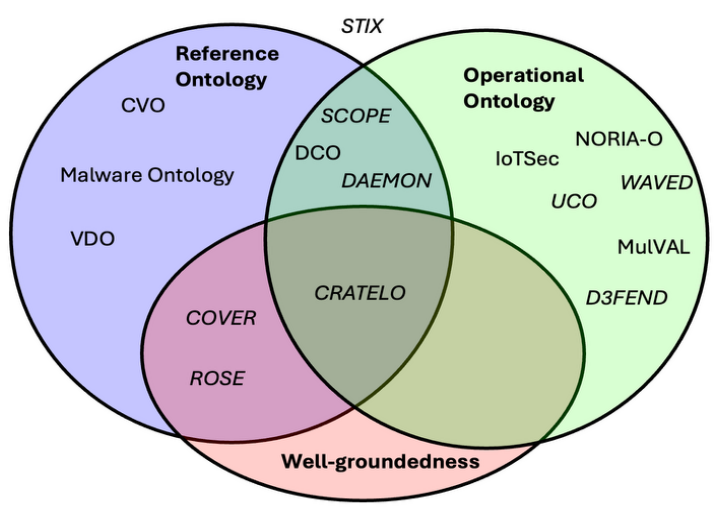

**Fig4**. Classification of cybersecurity ontologies across Reference, Operational, and Well-grounded criteria.

On the one hand, this exercise reveals that there are several ontologies of all ontologies close to what we are looking for except for being well-grounded as expected as only 7% are well-grounded. On the other hand, ontologies found doing this exercise highlight a crucial gap. Pure typological characterization (application versus domain, lightweight versus heavyweight) quickly tells us why an ontology looks right for a project, yet it says nothing about real-world usability. CRATELO, for instance, is structurally ideal but remains silent on standards or de facto industry standards such as MITRE ATT&CK. It is also no longer available in machine-readable form. Conversely, WAVED aligns perfectly with ATT&CK and D3FEND but lacks a foundational grounding and does not rely on a reference ontology, which may lead to logical contradictions and spurious inferences.

To be able to see WAVED as a possible candidate for our use-case, always by following a classification framework, we need to add more criteria to describe the ontology. Here, a criterion that would classify WAVED as valuable would be *credibility*, because WAVED relies on industry-validated de-facto standards. However, credibility is not a categorization criterion but a quality one. This highlights that, in addition to classification, evaluating ontologies is another key factor for selecting the most appropriate one for our use case. We can make the same remarks for STIX – which is an ontological standard –, UCO and D3FEND.

## 4 Ontologies evaluation criteria

### 4.1 Understanding Ontology Credibility: Concept, Dimensions and Limitations

We define credibility, adopted from the data quality standard: ISO/IEC 25012 [38], as the degree of confidence that users, developers, and experts place in it, particularly when applied to mission-critical domains like cybersecurity.

The definitions of Wilson et al. [39] for credibility rely on the notion of social quality from the work of McDaniel et al. [40]. Their approach decomposes credibility into three sub-dimensions:

- ***Authority*** reflects the degree to which an ontology is integrated within the broader semantic ecosystem, measured by the number of other ontologies that reused it and the extent of shared terminology across these connections.
- ***History*** captures the ontology's longevity and maintenance track record. It includes how long the ontology has been available as an open resource, how often it has been revised, and how frequently it has been used over time.
- ***Recognition*** aggregates download statistics, number of user reviews, and the positivity of those reviews.

Quantitatively, these dimensions can be computed through formulas that normalize each score against the best-in-class value across a reference set of ontologies producing a weighted credibility score. For instance, the Authority score is calculated as the ratio of links to the maximum number of links in the set; similar normalization applies for History and Recognition, each combining relevant variables like years of existence, number of revisions, downloads, and reviews. Together, these definitions reveal that credibility is not a vague or subjective trait but a measurable property, rooted in both social dynamics and technical interdependence. As ontologies become embedded in decision-making systems, credibility may well be the deciding factor between a theoretically perfect yet impractical ontology, and a well-used, battle-tested model that experts are willing to rely on. This calls for a systematic inclusion of credibility metrics in any ontology selection or evaluation process.

The first limitation of these credibility indicators lies in their practical measurability. Recognition demands not only download and review counts but a reliable method for separating positive from negative verdicts—a task easily confounded by balanced, academic prose and thus prone to subjective coding. Authority assumes a well-curated registry where ontologies inter-link according to shared conventions; in practice, such infrastructure is patchy, leaving link statistics sporadic and biased. History likewise hinges on complete version logs: beyond platforms like GitHub or BioPortal, records are often partial, forked, or missing, making cross-ontology comparisons laborious. Collectively, these data gaps undermine the robustness of any credibility score.

Beyond these measurement difficulties, Wilson et al. [39] also raises a more fundamental concern: credibility is context dependent. An ontology that was widely trusted a decade ago may have lost its relevance today, not because it was poorly designed, but because it fails to reflect the evolution of its domain. In cybersecurity, for instance, ontologies developed before 2013 could not incorporate de facto standards like MITRE ATT&CK, which is now a well-known reference. Even if those earlier models were authored by domain leaders, they cannot be considered credible today if they omit critical modern knowledge. Credibility, must therefore be also understood as context-dependent, tied not only to the ontology's internal features but to its temporal and situational relevance.

Yet, this does not mean we should avoid making credibility judgments. On the contrary, we need evaluation methods that acknowledge this temporal dependency and still enabling us to discriminate between outdated and current ontologies.

To address these challenges, we propose measurable indicators to evaluate easily the credibility criterion across periods and contexts.

### 4.2 Towards a new definition of cybersecurity ontology credibility

To make the credibility concept as a measurable and context-aware, we propose a decomposition into four dimensions, each reflecting the involvement of a specific stakeholder group in the ecosystem of the ontology's development. This structure acknowledges that credit in an ontology emerges from the validation of several sources: institutions, academia, day-to-day practitioners, and industry. Each dimension contributes a different form of legitimacy, and together they offer a well-rounded view of an ontology's credibility in practice. We define below each dimension, justify its relevance, and propose a practical evaluation indicator.

**Institutional Endorsement.** Institutions often act as gatekeepers in the development of security ontologies. They provide conceptual scaffolding: defining the key entities, relationships, and taxonomies that the ontology will encode. When an ontology is formally referenced by recognized standards (such as STIX, or NIST), it inherits a high level of legitimacy. The indicator here is deliberately binary: if the ontology is referenced by an established standard or de facto standard, it receives a score of 1; otherwise, it scores 0. While simplistic, this measure captures a powerful signal of trust and relevance at the institutional level.

**Academic Recognition.** The academic community plays a crucial role in ensuring that ontologies are grounded in rigorous methodology and up-to-date theoretical foundations. This form of credibility is assessed through traditional indicators: the quality of the publication venues (e.g., top conferences or journals in ontology engineering, cybersecurity, or AI). These measures reflect the extent to which the ontology is taken seriously by researchers,

**Day to Day Practitioners' Validation.** An ontology is considered day-to-day practitioners-validated when it has been reviewed by a panel of domain day-to-day practitioners, each meeting at least two of the following qualification criteria: advanced degree, at least 5 years of professional practice or active membership in a professional standards body.

**Industrial Maturity.** Nachabe and Jahan [41] suggest re-using the Technology Readiness Level (TRL) scale to gauge ontology maturity; TRL was created to model the technological maturation process of cyber-physical systems like a space shuttle or a satellite. The physical-related steps of this process do not apply to a pure software artefact like an ontology which can become operational in fewer steps. TRL also ignores maintenance steps and requires insider data to verify—a poor fit for an openly published, purely logical artefact. To keep the assessment practical, we introduce a streamlined, evidence-based indicator of industrial maturity that anyone can audit. We propose a three-level ladder:

- **L1 – Visibility:** at least one detailed production deployment is published—for example, a white paper, keynote speech, or deep-dive case study.
- **L2 – Documented deployment:** at least one use-case demonstrating the validity of the ontology has been published.
- **L3 – Independent adoption:** a panel of independent organizations reference the ontology in their technical artefacts (integration guides, architecture documents, RFPs, etc.).

Taken together, these four axes constitute a multi-perspective framework that captures credibility not as an abstract ideal, but as a concrete, measurable, and context-sensitive property. This evaluation method for credibility constitutes our answer to the second research question (**RQ2**).

## 5 New Ontologies Classification Framework Based on Credibility

Based on this new definition of credibility, we propose a **novel classification** of cybersecurity ontologies, organized into four distinct categories:

- **C1 – Academic Ontology**. The ontology has been published in at least a Q2-ranked journal or a B-tier conference.
- **C2 – Practical Ontology**. The ontology was co-authored by recognized domain day-to-day practitioners or developed with their involvement.
- **C3 – Standardized Ontology**. The ontology explicitly relies on established standards for the definition of all its core concepts.
- **C4 – Industrial Ontology**. The ontology has reached at least the second level of this criterion (L2).

An ontology may belong to multiple categories simultaneously. We applied our classification scheme to the 10 cybersecurity technologies highlighted in section 3.3 (see Figure 5).

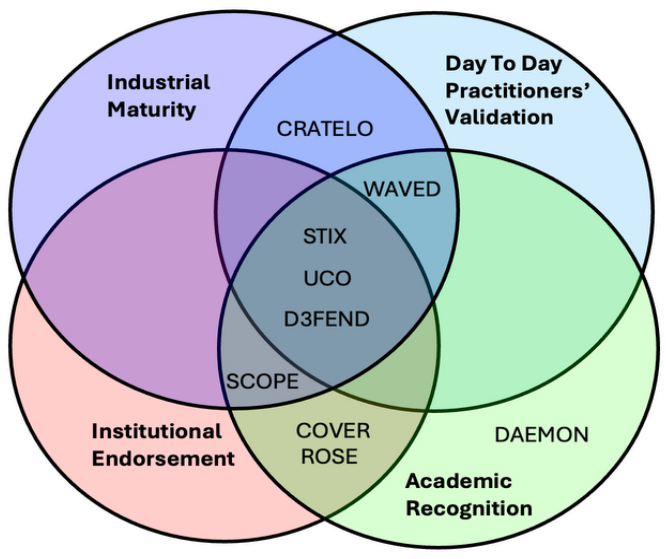

**Fig 5**. Classification of cybersecurity ontologies across Industrial Maturity, Expert Validation, Institutional Endorsement, and Academic Recognition.

Thanks to this framework, valuable ontologies that had slipped under the radar—such as WAVED and D3FEND—have now been brought into the spotlight, whereas the original framework largely left them on the sidelines. For instance, WAVED appears to be a very credible candidate due to its reliance on MITRE de facto standards (which are widely adopted in industrial context), on day-to-day practitioners' validation and peer reviewed. Compared to CRATELO which appears at first glance to be the perfect match we see that because of its lack of academic recognition and its lack of reliance on standards, it has become less credible. Nonetheless, some ontologies lack credibility in some dimensions but remain interesting because of their properties highlighted by F4OC, like COVER and ROSE, which are good candidate to be a reference ontology. These new classes allow us to consider more ontologies than F4OC alone, which are credible, and can result in more reutilization of ontologies.

## 6 A concrete use case: The ANCILE project

***ANCILE (AutoNomic Cyber-security with adversarIal Learning and Explanations***) is a Franco-Luxembourgish research project aiming at idealizing, prototyping and empirically evaluating, the next generation of SOAR frameworks. It proposes to architect them as two autonomic computing ***MAPE-K*** pipeline loops [42] in mirror image, one for cyber-defense and one for cyber-attack, allowing their co-improvement through adversarial machine learning in cyberattack-defense simulations.

As shown in the center of **Fig** , a MAPE-K loop manages a runtime reconfigurable system. **M** stands for a *Monitoring* service that senses data from both the system managed by the loop and the operations environment in which it is deployed, to identify relevant state changes in them. **A** stands for an *Analysis* service that receives relevant state change alerts from **M** and determines whether the current configuration of the system still satisfies its requirements in the new system and/or environment state. If this is not the case, the **M** triggers **P**, a *Planning* service that searches for a reconfiguration of the system that restores the satisfaction of those requirements. **P** then sends a reconfiguration plan to **E,** an *Execution* service that executes it. A MAPE-K loop is thus a continuous loop through a pipeline consisting of *Monitor, Analyze, Plan* and *Execute* steps. The **K** stands for a *Knowledge* base that is shared by the **M, A, P** and **E** pipeline services so that they can interoperate based on a common reference conceptual framework.

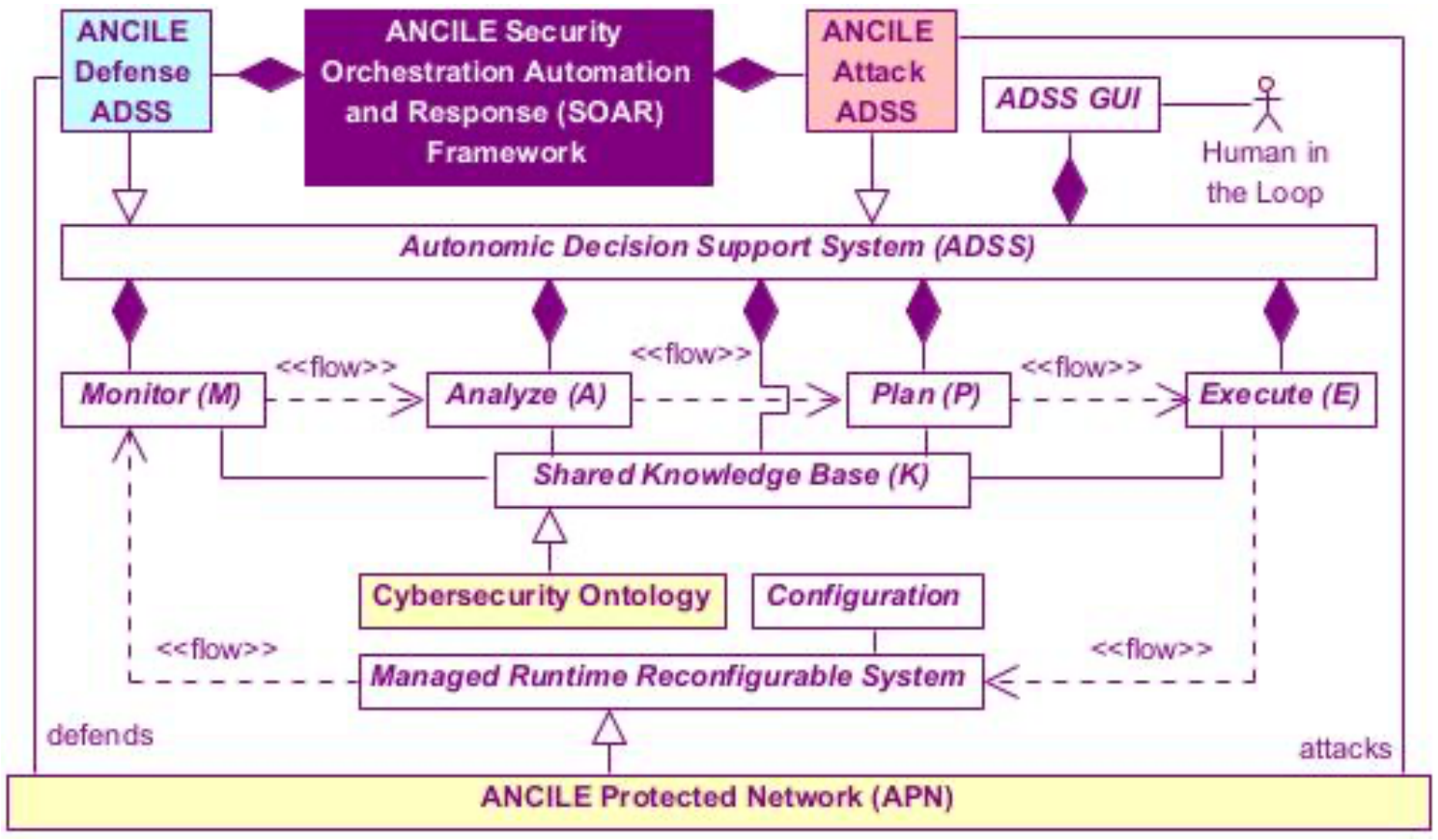

**Fig 6.** ANCILE Autonomic SOAR framework architecture

MAPE-K is an abstract autonomic computing architectural pattern that needs to be refined with concrete services specific to a particular application. Hence, in many autonomic systems, **K** takes the form of an application ontology, potentially re-using a domain ontology and/or a task ontology. Such ontology is in turn either queried by or imported inside the knowledge bases or models of the **M, A, P** and **E** inference services of the MAPE-K refinement. This is the case of the ANCILE SOAR framework architecture. in which the cyber-defense MAPE-K loop is an ***Autonomic Decision Support System (ADSS)*** [43], shown at the top-left of Figure 6, where the humans in the loop are (a) the blue team during adversarial training and (b) the SOC operators during cybersurveillance of a network. The cyber-attack MAPE-K loop is also an ADSS, shown at the top-right of Figure 6, of where the humans in the loop are the red team during adversarial training. In the ANCILE defense DSS, the alerts from **M** are cyber-attack alerts and the requirements are security quality criteria such as service availability and data confidentiality and integrity. They are collectively analyzed over time by **A** that tries to recognize a known multistep attack plan (*a.k.a.,* kill chain). The reconfiguration planned by **P** concerns the network protected by the ANCILE defense IDSS, one that satisfyingly minimizes the damage of a detected attack on these quality criteria. Such a plan may involve, for example, rerouting traffic from a host suspected to be compromised to another host provisioned to run the same service using an alternative technological stack that cannot be compromised through the same vulnerability. In ANCILE, **E** carries out such mitigation plan by using a ***Software Defined Network Controller (SDNC)*** [44] together with a container orchestrator such as Kubernetes [45].

Cybersecurity experts participating to ANCILE come from four distinct communities: (a) academic researchers as ESIEA[6], (b) cybersecurity tool developers, notably of the open-source OpenSOC SOC platform of ESIEA, and the GCAP-GCenter[7] comprehensive ***Network Detection and Response (NDR)*** solution commercialized by startup Gatewatcher, (c) SOC operators at ***SOC as a Service (SOCaaS)*** centers at both ESIEA and Gatewatcher and (d) cybersecurity standard specification contributors from ***LIST (Luxembourg Institute of Science and Technology)***, *notably.*ISO/IEC JTC 1/SC27. OpenSOC serves as the integration basis for the various services of the ANCILE SOAR. The GCAP-GCenter network attack signature and anomaly alert generators are key elements of the **M** service of ANCILE's defense ADSS. The SOC operators are the target users of ANCILE. Each of the **A**, **P** and **E** services of ANCILE's defense ADSS is the subject of one ongoing PhD. thesis research with radically innovative ideas with respect to the current state-of-the-art. However, the resulting prototype SOAR is intended to be industrialized into a commercial product after the project. The implementation platforms and reasoning engines of the ANCILE's **M**, **A**, **P** and **E** services are highly heterogeneous, ranging from deep neural networks in Python, to probabilistic logic programs in Prolog, to YAML specifications for Kubernetes and Java programs for the SDNC. *Therefore, the ANCILE ontology must satisfy the requirements from four distinct user communities while also remaining agnostic to implementation platforms.*

As explained earlier, as a **K** element in two MAPE-K loops, ANCILE's ontology must be an application ontology, reusing a domain ontology and multiple task ontologies since the tasks of the **M**, **A**, **P** and **E** elements in each loop are distinct (*e.g.,* attack detection *vs*. mitigation). Experts from the four communities listed above insist on the FAIR principles, the academic ones further advocate on foundational grounding, the tool developers prioritize credibility and standard contributors value official standard conformity.

As, no existing ontology simultaneously addresses all these concerns according to the new classification framework application in Figure 7, for ANCILE we have decided to integrate and align those that come close and complement each other to reach cover all of them. This ontology will be well-grounded in UFO and built upon existing efforts like ROSE and COVER. In terms of credibility, we aim to anchor our ontology in widely recognized standards such as ISO 31000 for risk modeling, and integrate established taxonomies like MITRE ATT&CK, CAPEC, CVE, along with ontologies such as D3FEND – all of which are component of the WAVED ontology – and STIX.

[6] ESIEA is the only higher education institution accredited by the French *National InformationSystem Security Agency (ANSSI)* to deliver the post-graduate diploma *"Information System Security Expert"*.

[7] GCAP-GCenter is one of only four NDR solutions that are both certified and qualified by ANSSI as adhering to common criteria and providing sufficient detection capabilities.

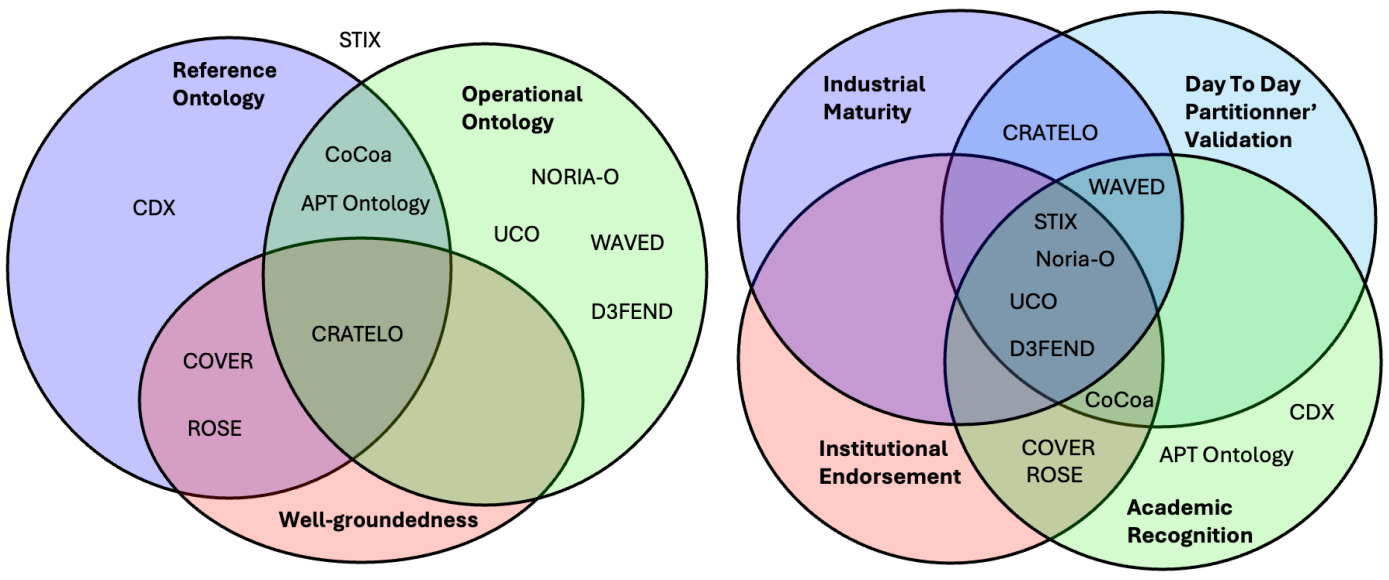

**Fig 7.** Application of the complete framework for the ANCILE use-case

CRATELO is not considered because the original artefact appears to be lost. Moreover, our ontology is being developed in collaboration with cybersecurity experts of diversified professional roles: educators, researchers, developers of SOC platforms and commercial NDR tools, SOC operators and standard contributors. It will be deployed in a proof-of-concept, cloning the functionalities of a real-world web application on a different network featuring cutting-edge automation technology, — corresponding to a level 2 of maturity according to our criterion. Through this process, we aim to produce not only a technically sound but also a credible ontology, ready to support high-stakes reasoning in autonomous cybersecurity systems.

## 7 CONCLUSION

This study used the F4OC framework to analyze 87 cybersecurity ontologies, revealing most lack foundational grounding and remain tied to narrow, use-case-specific contexts limiting reuse (RQ1). Structure alone being insufficient, we refined Wilson & McDaniel's notion of "credibility" into four measurable indicators — institutional, academic, expert and industrial — forming a practical evaluation scheme (RQ2). Pairing these indicators with F4OC yielded four credibility-based classes (academic, expert, standardized, industrial), offering a richer lens for ontology selection.

Applying the scheme to the ANCILE SOAR project showed that no single ontology met all requirements, yet several scored highly enough to act as robust starting points for a bespoke model. The work therefore offers a screening tool for practitioners and a roadmap for developers aiming to raise the trust profile of their ontologies.

Key limitations remain: our literature review was not fully systematic, and the history dimension of credibility—stability and evolution—was only partially captured. Future research should address these gaps and extend the evaluation to the full set of ontology quality criteria. Future work needs also to evaluate this credibility criterion with cybersecurity experts and to look for its application to other domains.

**Acknowledgements.** This work was supported by the French National Research Agency (ANR) under the ANCILE project (grant number ANR-23-CE39-0010).

**Disclosure of Interests**. The authors have no competing interests to declare that are relevant to the content of this article.